\documentclass[%
 reprint,
 amsmath,amssymb,
 aps,
]{revtex4-2}

\usepackage{graphicx}
\usepackage{dcolumn}
\usepackage{bm}
\usepackage{hyperref}
\usepackage[normalem]{ulem}
\hypersetup{colorlinks=true, citecolor=blue, urlcolor=blue, linkcolor=blue}

\begin{document}

\preprint{APS/123-QED}

\title{Fast Penning Ionization of Cold Rydberg atoms in an Electric Field}

\author{Changjie Luo$^{1,3}$}%
\author{Feng Fang$^{1,2}$}%
 \email{fangfeng@gdlhz.ac.cn}
\author{Wenchang Zhou$^{1}$}%
\author{Peng Zhang$^{4}$}%
\author{Xinwen Ma$^{1}$}%
\author{Jie Yang$^{1}$}%
 \email{jie.yang@impcas.ac.cn}
 
\affiliation{$^1$Institute of Modern Physics, Chinese Academy of Sciences, Lanzhou 730000, China}

\affiliation{$^2$Advanced Energy Science and Technology Guangdong Laboratory, Huizhou 516000, China}

\affiliation{$^3$School of Physical Sciences, University of Chinese Academy of Sciences, Beijing 100049, China}

\affiliation{$^4$Department of Physics, Renmin University of China, Beijing 100872, China}




\date{\today}

\begin{abstract}
We observe a fast Penning ionization in a dilute gas of cold rubidium Rydberg atoms, in the presence of a static electric field of 50 V/cm,
with the ionization rate coefficients for two specific states being measured as $\widetilde{\Gamma}_{51s}=3.8(9)\times10^7$ Hz$\cdot$($\mu$m)$^6$ and $\widetilde{\Gamma}_{49d}=1.2(4)\times10^8$ Hz$\cdot$($\mu$m)$^6$,
which are $3\sim8$ orders of magnitude higher than the theoretical predictions in field-free space.
Our analysis based on a polarized two-atom model reveals that
the ionization threshold of Rydberg atoms is lowered by the static electric field, reducing the energy exchange required for Penning ionization and increasing the ionization rate. 
Beyond this, the dipole-dipole interaction strengthened by the electric field between two Rydberg atoms at a micrometer-scale distance leads to double ionization of the atoms pair, opening a new autoionization channel.
Such enhancement of the Penning ionization by a static electric field poses both a threat to the stability and a potential control strategy for quantum systems composed of cold Rydberg atoms with micrometer-scale interatomic separations.

\end{abstract}

\maketitle


{\it $Introduction$} The cold Rydberg gas exhibits strong interatomic and many-body interactions~\cite{PhysRevLett.95.253002,PhysRevLett.97.083003,PhysRevLett.100.043002}, making it an ideal platform for quantum information processing~\cite{PhysRevLett.85.2208,RevModPhys.82.2313,10.1116/5.0036562} and quantum simulation~\cite{PhysRevLett.124.193401,PhysRevLett.131.083601,PhysRevX.8.021070,PhysRevX.11.031005}. 
In optical lattices or tweezer arrays, the spacing of cold Rydberg atoms can be precisely tuned from hundreds of nanometers to tens of micrometers, where interatomic interactions are typically governed by dipole–dipole or van der Vaals interactions~\cite{PhysRevA.88.043436,PhysRevLett.124.103601,PhysRevLett.131.203003}.
Notably, while strong dipole–dipole interactions(DDI) enable coherent dynamics, they can also trigger non-radiative energy transfer processes, such as Penning ionization or state redistribution. These processes constitute significant sources of decoherence in such quantum systems~\cite{RevModPhys.82.2313,10.1116/5.0036562}.

Penning ionization is the reaction
\begin{equation}
{\rm A}^\ast + {\rm B} \rightarrow {\rm A} + {\rm B}^+ + {\rm e}^-,
\label{eq:1}
\end{equation}
where ${\rm A}^\ast$(${\rm A}$) is an atom or molecule in an excited state such as Rydberg state (in its ground or low-lying excited state).
The partner ${\rm B}$ can range from a single atom to a complex system such as a macroscopic metal crystal with adsorbed molecules on its surface, and may be in a ground or excited state~\cite{RevModPhys.65.337}.
In the reaction products, ${\rm B}^+$ represents the ionized form of ${\rm B}$, and ${\rm e}^-$ denotes the emitted electron.
Since its first observation in neon and argon systems by F. M. Penning in 1927~\cite{Penning1927}, Penning ionization has attracted considerable interest across a variety of physical systems and has been linked to several important decay processes.
In particular, for Rydberg–Rydberg atomic collisions at high collisional energies~\cite{10.1063/1.1673523,PhysRevLett.43.126,R.E.Olson_1979,PhysRevA.46.1132,PhysRevLett.61.2088,PhysRevA.46.5795,Robicheaux_2005}, both experimental and theoretical studies have demonstrated that the Penning ionization cross section closely approximates the geometric cross section, scaling as the fourth power of the principal quantum number, $n$~\cite{R.E.Olson_1979}. 
Furthermore, the cross section reaches maximum when the collision velocity $v \approx 1.25v_e$, where $v_e$ denotes the velocity of the valence electron~\cite{PhysRevLett.43.126,R.E.Olson_1979}. 
Moreover, although  Penning ionization in cold Rydberg atom systems received limited attention, both theoretical~\cite{Robicheaux_2005,WOS:000266566600008,Kiffner_2016,Efimov_2016,PhysRevA.102.032819} and experimental efforts~\cite{PhysRevA.73.020704,PhysRevLett.124.253201} have begun to explore this phenomenon.


In this work, we investigated Penning ionization in cold Rydberg atoms with interatomic separations on the micrometer scale—an interaction regime previously considered inapplicable for such processes.
We observed rapid ionization of Rydberg atoms in the presence of an external static electric field. Remarkably, the measured ionization rates are 3 to 8 orders of magnitude higher than theoretical predictions made under field-free conditions~\cite{WOS:000266566600008,Kiffner_2016,PhysRevLett.124.253201}.

Such a significant enhancement effect can be explained by the following two mechanisms:  
(1) Reduction of Ionization Threshold: The applied static electric field lowers the ionization threshold of Rydberg atoms, reducing the energy exchange required for Penning ionization and thereby increasing the probability of ionization.
(2) Strengthened Dipole-Dipole Interactions (DDI): The static electric field induces polarization in individual Rydberg atoms, rendering the effect of DDI between two Rydberg atoms significant.
In spatial regions where the DDI is attractive, the ionization threshold is effectively lowered, leading to the simultaneous ionization of both atoms. In other words, the  static electric field together with the  DDI induce a novel Penning ionization process:
 \begin{equation}
{\rm A}^\ast + {\rm A}^\ast \xrightarrow{\text{static electric field}} {\rm A}^+ + {\rm A}^+ + {\rm e}^-+ {\rm e}^-,
\label{nim}
\end{equation}
which is very different from the one shown in Eq.~(\ref{eq:1}), and significantly enhance the ionization of Rydberg atoms.

Our results not only reveal a new Penning ionization mechanism, but also indicate that, in cold Rydberg atom systems, Penning ionization may lead to non-negligible atom loss when a static electric field is applied. 
This process has the potential to compromise system stability, while also offering a means for controlled manipulation.

\begin{figure}
\includegraphics[width=\columnwidth]{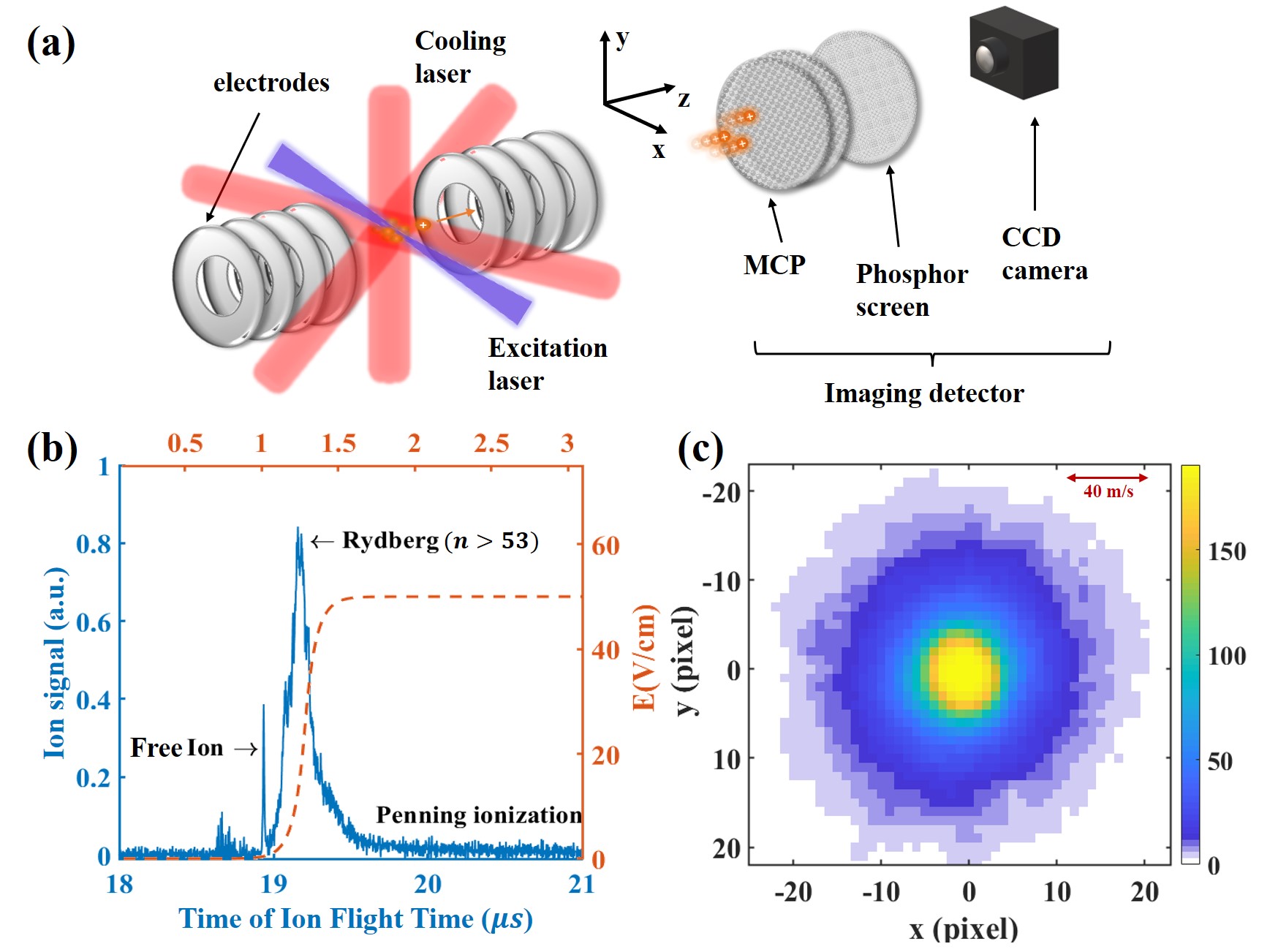}
\caption{\label{fig:1}{\bf (a)} Diagrammatic sketch of the experimental setup, the compositions of MOT and ion spectrometer. {\bf (b)} The extraction pulsed electric field (dash line) and the ionic ToF spectrum (solid line) for the initial 49d Rydberg atoms with density of $1\times10^8\ {\rm cm}^{-3}$. The atoms are excited by the pulsed laser at t=0, and the pulsed electric field is applied 1 $\mu$s later, shown by the labels on the top and right. The time interval between the laser and the electric field is minimized as much as possible, preventing the loss and redistribution of initial Rydberg atoms (less than $5\%$). The sharp peak and the broad peak in TOF spectrum represent the free ions and the ions from field-ionization of high-n Rydberg atoms, respectively. {\bf (c)} The image recorded by VMI, arising from the ions in the interval [19.6-20.6 $\mu$s] in TOF. The image was accumulated for 1000 repeat rimes.The two-dimensional velocity distribution of ions can be derived through the calibrated scale from space to velocity.}
\end{figure}

{\it Experimental Setup and Process} The experimental setup was thoroughly introduced in our previous work~\cite{10.1063/5.0033595}.
As shown in Fig.~\ref{fig:1}(a), a standard magneto optical trap (MOT) was used, and up to $1\times10^7$ $^{87}$Rb atoms were cooled to 200 $\mu$K and trapped within a region of 0.7 mm~\cite{10.1063/5.0033595,LI201923}. 
A pulsed dye laser (duration 5 ns, wavenumber 20826.5 cm$^{-1}$ or 20826.1 cm$^{-1}$, line-width 0.02 cm$^{-1}$, focused beam waist size 350 $\mu$m) excited the cold atoms from $\mathrm{5^{2}P_{3/2}}$ state to Rydberg states (either $\mathrm{51^{2}S_{1/2}}$ or $\mathrm{49^{2}D_{5/2}}$). 
The cooling lasers were switched off 50 ns after the dye laser pulse to allow most 5P state atoms to decay back to the 5S ground state~\cite{PhysRevA.57.2448}, minimizing perturbations to the evolution of Rydberg atoms. 

The ion spectrometer, which is composed of eight pairs of circular electrodes symmetrically positioned along the z-axis and an imaging detector, is used to detect the ions generated by ionization of Rydberg atoms. 
The imaging detector consists of two microchannel plates (MCP), a phosphor screen, and a CCD camera, and has a high sensitivity of single-ion detection~\cite{ZHOU2021116516,10.1063/5.0120819}. 
In some cases, the imaging detector can work in the “gating mode” in which a pulsed high voltage applies on the MCP, offering a minimal exposure time of 50 ns. 
The electric field is specially designed to realize the function of velocity map imaging (VMI), by which ions with the same velocity in source are mapped onto the same spatial position on detector, regardless of their initial spatial position. Imaging can provide information on the number and velocity distribution of product ions with a signal-to-noise ratio as high as single ion detection.
The ion spectrometer provides two kinds of signal, i.e., the time-of-flight (ToF) spectra and the image of VMI.
A pulsed electric field having a rising stage of 400 ns and a maximum magnitude of 50 V/cm in the reaction region was applied by the electrodes with a delay of 500 ns after the excitation laser. 
The strength of electric field ensures the field-ionization of Rydberg atoms $n\ge54$, but is lower than the field ionization limit of $n=50$ state, which is about 67 V/cm~\cite{F.Merkt_1998}.

Since the laser excitation, a small fraction of Rydberg atoms was ionized during the spontaneous evolution, generating free ions and electrons~\cite{PhysRevLett.85.4466,PhysRevA.69.063405}. 
Meanwhile, some Rydberg atoms underwent state redistribution through inelastic collisions with free electrons or other Rydberg atoms~\cite{PhysRevA.69.063405,PhysRevLett.100.123007}. When the pulsed electric field was applied, the free ions were extracted toward the detector, and the corresponding signal manifested a sharp peak at 18.9 $\mu$s in ToF spectrum as shown in Fig.~\ref{fig:1}(b). 
As the electric field strength gradually increased to the maximum, the high-n Rydberg atoms ($n \ge 54$) underwent field-ionization, producing the broad peak from 19 to 19.5 $\mu$s in ToF spectrum.

As the electric field reached the maximum magnitude, the free electrons and ions were removed, and only the Rydberg atoms were left in reaction region. 
Since then, the product ions mainly came from Penning ionization, with small contributions from the ionization induced by blackbody radiation (BBR)~\cite{PhysRevA.26.1490,PhysRevA.75.052720,PhysRevA.70.042713}. Other processes, such as many-body ionization~\cite{PhysRevLett.100.043002} and tunneling ionization~\cite{PhysRevA.99.013421}, were negligible under our experimental conditions. 
The number of ions was counted from the image of VMI. 
The imaging detector worked in “gating mode” with exposure time of 1 $\mu$s, corresponding to the time interval of 19.6 to 20.6 $\mu$s in ToF srectra, to prevent the interatomic distance change caused by atomic motion. 
A typical image is shown in Fig.~\ref{fig:1}(c).
For each experimental exposure, the ion count ranged from 2 to 100 on average. We repeat the exposure 1000 times for each experimental condition to derive a reliable counting of ions. 

As a result, the ionization rate $\gamma_{exp}$ is calculated by
\begin{equation}
\gamma_{exp}=N_i/(N_R\times\Delta t) ,
\end{equation}
where $N_i$ is the number of ions counted from the image, $\Delta t$ is the reaction time of 1 $\mu$s, corresponding to the exposure time, and $N_R$ is the number of Rydberg atoms. 
To estimate $N_R$, the magnitude of electric field was set to 70 V/cm, the initial Rydberg atoms in 49d or 51s state are all field-ionized~\cite{F.Merkt_1998}. The ToF spectrum provided the value of $N_R$ with an error of 10$\%$, which also represents the error of $\gamma_{exp}$ as the dominated source. By averaging results over 1000 repetitions, we obtained the $\gamma_{exp}$, listed in TABLE~\ref{tab:1}, with the Rydberg state and density adjusted by varying the excitation laser's wavenumber and intensity.

\begin{table*}
\caption{\label{tab:1}the total ionization rates for 49d and 51s state}
\begin{ruledtabular}
\scriptsize
\begin{tabular}{ccccc|cccccc}
&\multicolumn{4}{c}{\bf49d}&\multicolumn{6}{c}{\bf51s}\\
&\bf{exp.1}&\bf{exp.2}&\bf{exp.3}&\bf{exp.4}&\bf{exp.1}&\bf{exp.2}&\bf{exp.3}&\bf{exp.4}&\bf{exp.5}&\bf{exp.6}\\
\hline
$N_R$&$1.8(1)\times10^3$&$3.3(1)\times10^3$&$4.9(2)\times10^3$&$7.3(3)\times10^3$&$1.1(1)\times10^3$&$2.2(1)\times10^3$&$7.3(3)\times10^3$&$9.5(4)\times10^3$&$1.0(1)\times10^4$&$1.2(1)\times10^4$\\
$N_i$&$7.5(3)\times10^0$&$2.4(1)\times10^1$&$4.9(2)\times10^1$&$1.0(1)\times10^2$&$2.1(1)\times10^0$&$7.1(3)\times10^0$&$3.4(1)\times10^1$&$6.2(2)\times10^1$&$7.2(3)\times10^1$&$1.2(1)\times10^2$\\
$\gamma_e$(Hz)&$4.2(2)\times10^3$&$7.1(4)\times10^3$&$1.0(1)\times10^4$&$1.4(1)\times10^4$&$1.9(1)\times10^3$&$3.2(2)\times10^3$&$4.6(3)\times10^3$&$6.5(4)\times10^3$&$7.1(4)\times10^3$&$9.5(4)\times10^3$
\end{tabular}
\end{ruledtabular}
\end{table*}

{\it Results and Analyses} The average thermal velocity of cold atoms is lower than 0.2 m/s, indicating that the atomic motion during the reaction time is less than 0.2 $\mu$m. 
So, the reaction rate, instead of the cross-section, becomes an appropriate parameter for describing the reaction between cold particles. The average distance between the Rydberg atoms in experiment is in the range of 16 to 34 $\mu$m. 
It is 2 orders of magnitude larger than the diameter of valence electron orbital, for example, 100 nm for a Rydberg atom with n = 50. Under this condition, the Penning ionization rate $\Gamma$ for a pair of Rydberg atoms is inversely proportional to the sixth power of the interatomic distance $r$~\cite{WOS:000266566600008}
\begin{equation}
\Gamma_{r}=\widetilde{\Gamma}_{nl}/r^6 ,
\end{equation}
where $\widetilde{\Gamma}_{nl}$ is the coefficient for the atoms in $nl$ Rydberg state. The ionization rate decreases dramatically with the increasing of $r$, meaning that only the Penning ionization induced by the nearest neighboring atom needs to be considered in a dilute Rydberg gas. 
Thus the ionization rate $\gamma_{N}$ for a Rydberg gas of  $N$ atoms can be expressed as:
\begin{equation}
\gamma_{N}=\frac{1}{N}\int{f_{N}(r)\cdot \Gamma_{r}\cdot dr} ,
\label{eq:4}
\end{equation}
where $f_N(r)$ represents the distribution of the shortest distance between any Rydberg atom and its neighbor atoms, and it is accessible through a Monte Carlo numerical simulation.

\begin{figure}
\includegraphics[width=\columnwidth]{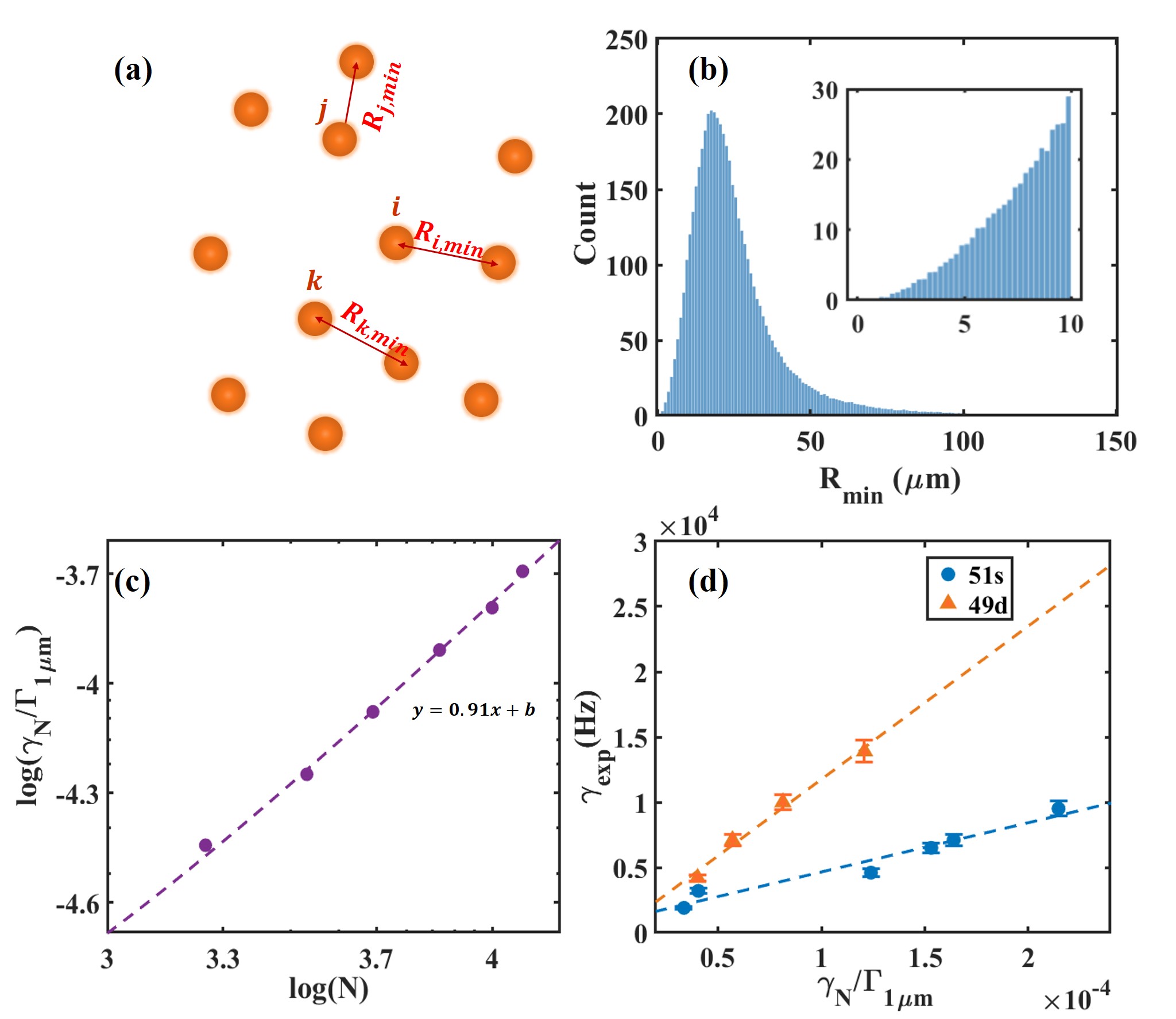}
\caption{\label{fig:2}{\bf (a)} For Rydberg atom i, the shortest interval is denoted as $R_{i,min}$. {\bf (b)} The shortest interval distribution within the range of 0 to 150 $\mu$m in case of N=4900, with a bin width of 1 $\mu$m. The inset shows the shorter range of 0 to 10 $\mu$m, with a bin width of 0.25 $\mu$m. {\bf (c)} The rate $\gamma_{penning}(N)$ normalized to $\gamma_{1\mu m}$. Circles represent the numerical simulation results, while the dashed line corresponds to the fitted line following a 0.91 power law relationship. {\bf (d)} The plot of experimental average rates $\gamma_{exp}$ vs simulated relative Penning ionization rates. Experimental measurement results are represented by circles, while the dashed line depicts the fitted line.}
\end{figure}

Fig.~\ref{fig:2}(b) shows an example of the simulated $f_N(r)$ for the Rydberg gas following a three-dimensional Gaussian distribution with $\sigma=175~\mu{\rm m}$ and $N=4900$. 
The simulation repeated 100 times for statistical reliability. Fig.~\ref{fig:2}(c) plots the parameter $\gamma_{N}/\Gamma_{1\mu m}$ versus $N$, and the fit reveals that $\gamma_{N}/\Gamma_{1\mu m}$ scales with $N^{0.91}$, exactly matching the experimental observation~\cite{PhysRevA.73.020704}. 
Finally, the measured $\gamma_{exp}$ was plotted as a function of $\gamma_{N}/\Gamma_{1\mu m}$, as shown in Fig.~\ref{fig:2}(d). By fitting the linear relationship, the Penning ionization rate coefficients are derived as ${\widetilde{\Gamma}}_{51s}=3.8(9)\times{10}^7\ {\rm Hz}\cdot{(\mu {\rm m})}^6$ and ${\widetilde{\Gamma}}_{49d}=1.2(4)\times{10}^8\ {\rm Hz}\cdot(\mu {\rm m})^6$. 
Apparently, the ${\widetilde{\Gamma}}_{49d}$ is larger than ${\widetilde{\Gamma}}_{51s}$.
The polarizability of 49d Rydberg state is 1.6 times larger than that of 51s state~\cite{PhysRevA.98.052503,Bai_2020}. In electric field, the larger polarizability leads to a stronger DDI and subsequently the higher Penning ionization rate. 
More importantly, the experimental result indicates that the Penning ionization rate is on the order of 100 MHz for $n\approx50$ Rydberg atoms at a separation of 1 $\mu$m in a 50 V/cm electric field. It is $3\sim8$ orders of magnitude higher than the theoretical predictions under field-free condition. 

In 2016, M.Kiffner $et$ $al$ predicted that the Penning ionization rate increases dramatically when the wave functions of two Rydberg atoms overlap~\cite{Kiffner_2016}. 
Soon it was experimentally observed by M. Mizoguchi $et$ $al$~\cite{PhysRevLett.124.253201}, and they found that around 50$\%$ cold Rydberg atoms ($n=57$) in an optical lattice with interatomic distance of 532 nm possess Penning ionization in 60 ns.
It should be emphasized that the dilute cold Rydberg gas has a micrometer-scale interatomic distance in our experiment, which is significantly larger than the size of valence electron orbitals. 
This suggests that there are some different enhancement mechanisms in which the electric field may have a significant influence on the Penning process.

{\it Reduction of Ionization Threshold}  For an atom in the presence of an electric field, there will be a saddle point in the potential of the electron. The atom will be ionized classically if the energy of the atom lies above the saddle point.
So, a lower saddle point indicates that a smaller amount of energy exchange is required for the Penning ionization.
As a simple estimation, the electric field of 50 V/cm decreases the ionization energy of Rb atoms to 5.5 $cm^{-1}$ above the binding energy of 51s or 49d, which is equivalent to the ionization energy of $n\approx150$ Rydberg state in field-free space. 
Considering that the Penning ionization rate is proportional to $n^{5.5}$ of the principal quantum number~\cite{WOS:000266566600008}, the rate of $n=150$ Rydberg atom is expected to be approximately 4000 times greater than that of atoms with $n=50$. 
However, these estimations are still insufficient to explain the reason for such an increase in ionization rate, which indicates the interactions between Rydberg atoms may further enhance the Penning process in the presence of an electric field.

{\it Novel Penning Ionization Process} Specifically, in the presence of the static electric field, atoms tend to be polarized along this field, rather than randomly. 
As a result, the average effect of the interatomic DDI becomes non-negligible. Explicitly, it is proportional to $(1-3\cos^2\theta)/R^3$, where $\theta$ is the angle 
between the relative position $\bm{R}$ of two atoms and the electric field, and
$R=|{\bm R}|$.
Thus, in the region with $\theta<\arccos(1/\sqrt{3})\approx{0.39}\pi$, the DDI is attractive and reduces the potential energy experienced by the electrons. 
When the two atoms are close enough so that this potential energy is reduced below the energy of the Rydberg state, the electrons of both of the two atoms can escape simultaneously. I.e., the double ionization process shown in Eq.~(\ref{nim})
can occur.

\begin{figure}
\includegraphics[width=\columnwidth]{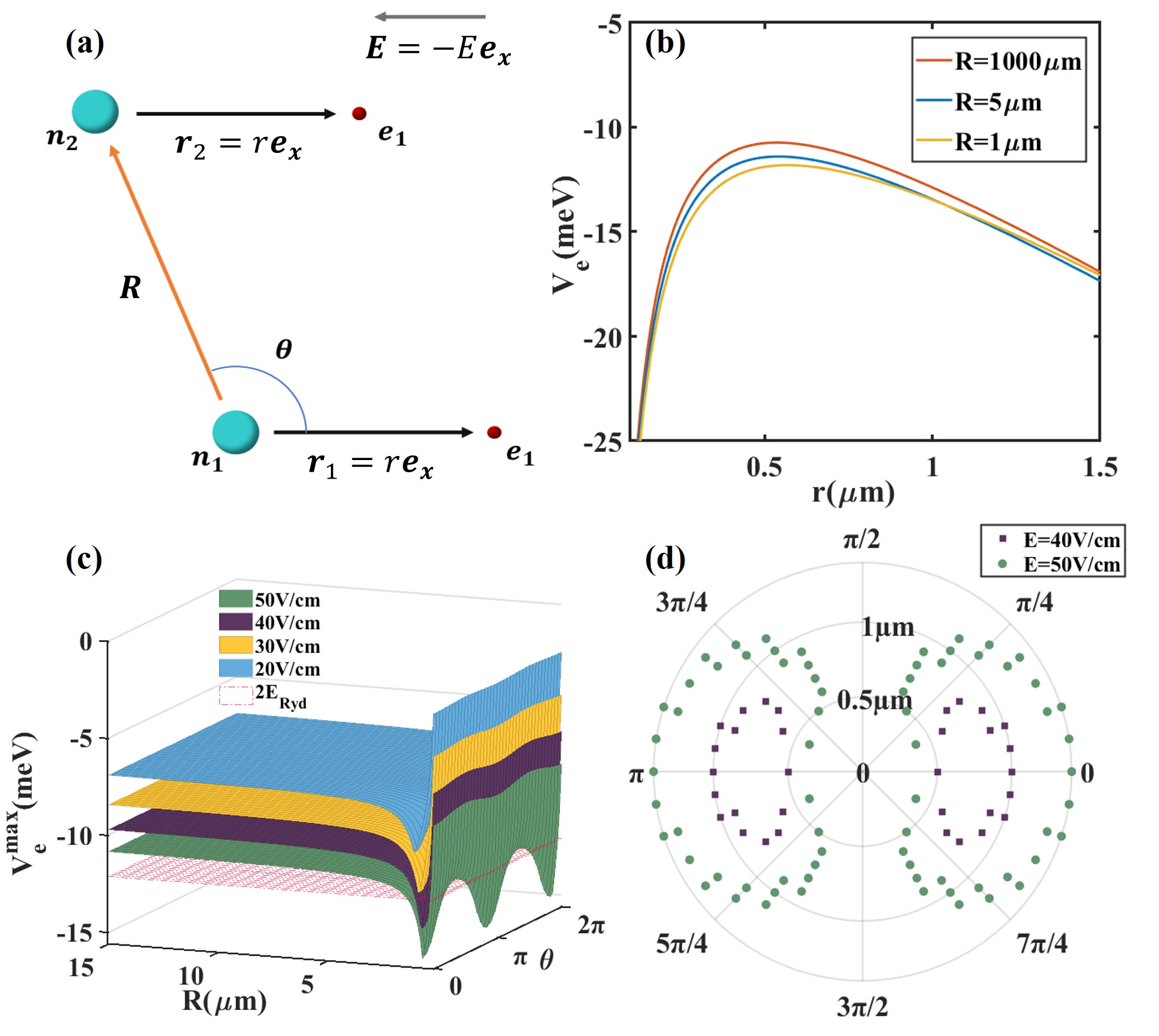}
\caption{\label{fig:3}{\bf (a)} The electrons ($e_1$, $e_2$) and nuclei ($n_1$, $n_2$) of two Rydberg
atoms 1 and 2 in a static electric field. {\bf (b)} The electronic potential energy $V_e$ as a function of electron-nucleus distance $r$, for $E=$ 50V/cm, $\theta = \pi/2$ and various values of $R$. {\bf (c)} The maximum of potential energy $V_e^{\rm max}$ as a function of $R$ and $\theta$, for various electric field strength $E$. The red plane indicates $2E_{
\rm Ryd}$ for two $^{87}$Rb atoms in the 51$s$ Rydberg state. {\bf (d)} A plot in polar coordinates $(R,\theta)$. The condition (\ref{con}) is satisfied for two $^{87}$Rb atoms in the 51$s$ Rydberg state in the areas enclosed by the points.}
\end{figure}

This preliminary physical picture can be further verified with a semi-quantitative estimation. As shown in Fig.~\ref{fig:3}(a), we consider two Rydberg atoms 1 and 2, which are both polarized by the static electric field. 
The electron-nuclear relative position ${\bm r_i}$ of the atom i (i=1,2) align in antiparallel to ${\bm E}$. The potential energy of the two electrons in this system can be calculated as:
\begin{eqnarray}
      V_e =&&\frac{1}{4\pi\epsilon_0}\left(-\frac{e^2}{|{\bm r_1}|}-\frac{e^2}{|{\bm r_2}|}-\frac{e^2}{|{\bm R}-{\bm r_1}|}-\frac{e^2}{|{\bm R}+{\bm r_2}|}\right.\nonumber\\
      &&\left.+\frac{e^2}{|{\bm R}+{\bm r_2}-{\bm r_1}|}\right)+e{\bm E}\cdot({\bm r_1}+{\bm r_2}) ,
      \label{eq:6}
\end{eqnarray}
with ${\bm R}$  being the relative positions of the two atoms, and $e$ and $\epsilon_0$ being the electronic charge and vacuum dielectric constant, respectively. 
Note that, for the sake of rigor, here we do not adopt the aformentioned effective form of the DDI. 
Instead, we consider all Coulomb interactions between charged particles from first principles.
Specifically, the first term in the right-hand side of Eq.~(\ref{eq:6}) represents the Coulomb potential energy between electrons and nuclei, as well as the interaction between electrons, while the second term represents the potential energy of the two electrons in static electric field. Moreover, without loss of generality, here we assume the static electric field is along the $-{x}$ direction, i.e., ${\bm E}=-E{\bf e}_x\  (E>0)$. Additionally, since the atoms are influenced equally by the electric field, we take ${\bm r_1}={\bm r_2}=r{\bf e}_x$, where $r$ is the distance between the electron and the relevant nuclear. 
It is clear that the electronic potential energy $V_e$ of Eq.~(\ref{eq:6}) can be considered as a function of the electron-nuclear distance $r$, with the following parameters: the interatomic distance $R=|{\bm R}|$, the angle $\theta$ between ${\bm R}$ and $x$-axis, and the electric field strength $E$.  As examples, in Fig.~\ref{fig:3}(b) we show $V_e(r)$ for $\theta=\pi/2$, $E=50$V/cm, and various interatomic distance $R$.

Furthermore, the maximum value of $V_e(r)$ is a function of the parameters $(R,\theta,E)$, and can be denoted as $V_e^{\rm max}(R,\theta,E)$. We suppose that the two electrons can escape from the corresponding nuclei (i.e., the process of (\ref{nim}) can occur) under the condition
\begin{eqnarray}
 V_e^{\rm max}(R,\theta,E)<2E_{\rm Ryd},\label{con}
 \end{eqnarray}
with $E_{\rm Ryd}$ being the binding energy of a single atom. Now the question is: can the static electric field make this condition to be satisfied?  
  
To answer this question, in Fig.~\ref{fig:3}(c) we plot $V_e^{\rm max}$ as functions of $(R,\theta)$, for $E=$ 20$\sim$50 V/cm. It is shown that $V_e^{\rm max}$ continually decreases as the static electric field strength $E$ increases, and thus the condition (\ref{con}) can be satisfied in some regions of $(R,\theta)$.
In Fig.~\ref{fig:3}(d) we further illustrate these regions 
for two atoms in 51$s$ state, with $E=40$ and $50$ V/cm. For interatomic distance $R\sim1.5\ \mu$m, the process (\ref{nim}) occurs over a relatively wide angular range. 

Furthermore, the product ions from double ionization channel are examined by VMI technology. 
We utilized velocity map imaging (VMI) techniques to measure the velocity of the product ions (details are provided in the Supplementary Material).
Taking Fig.~\ref{fig:1}(c) as an example, most ions are concentrated within a region no wider than 10 pixels, while a small fraction of ions disperses in a large surrounding area. 
The corresponding velocity even exceeds 40 m/s, far greater than the thermal velocity of Rydberg atoms (0.2 m/s). 
For comparison, the maximum velocity achievable from Coulomb repulsion between two Rb ions at an internuclear distance of 1 $\mu$m is 40 m/s. This suggests that such substantial kinetic energy must originates from strong Coulomb repulsion between ion-pair, thereby providing evidence of the double ionization process in Rydberg atom-pairs.

\begin{figure}
\includegraphics[width=\columnwidth]{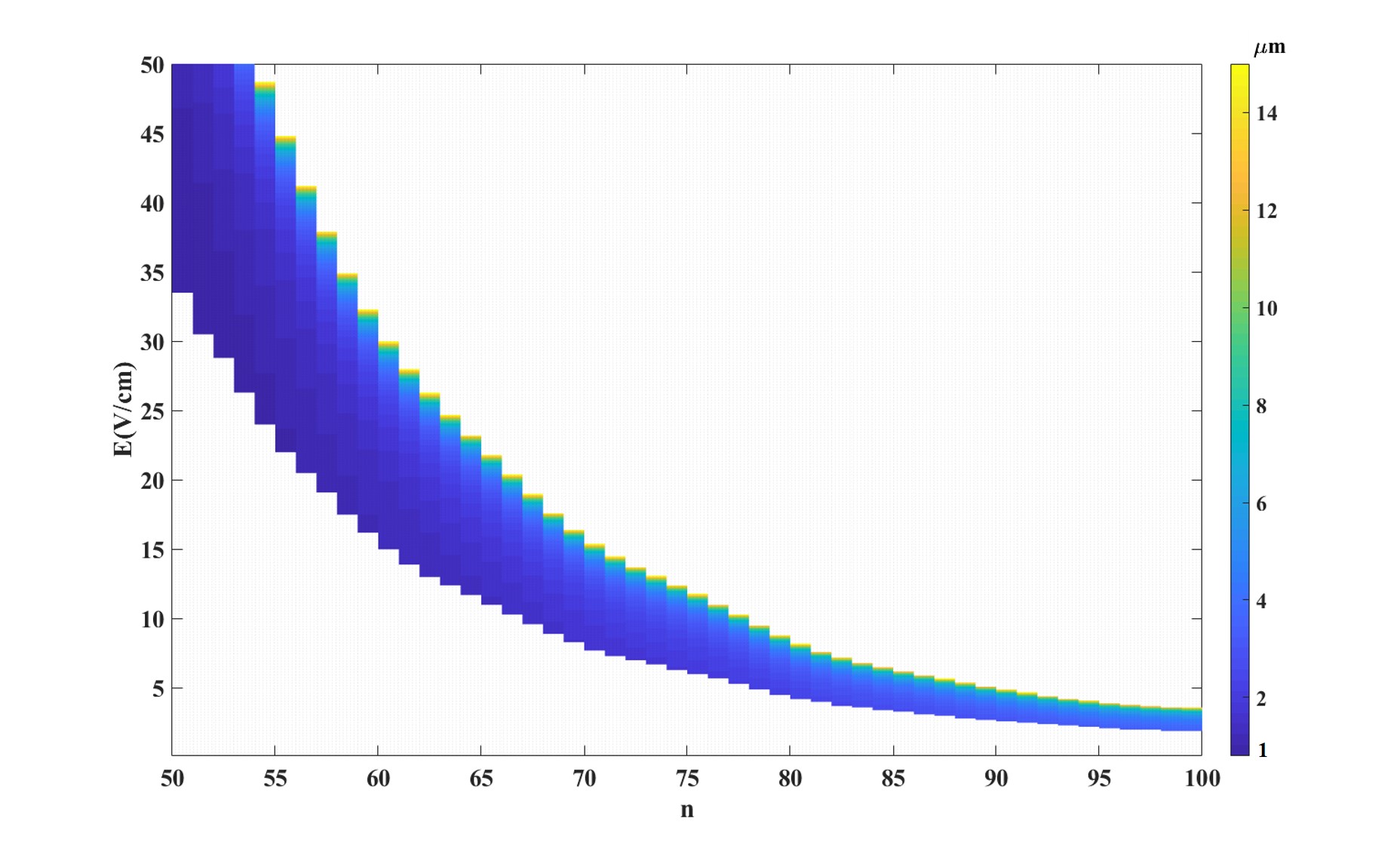}
\caption{\label{fig:4}The threshold interatomic distance as the functions of Rydberg state (principal quantum number n) and electric field strength (E in V/cm). 
The interatomic distance is represented by different colors in this figure. A necessary condition for double ionization of a Rydberg atom pair is that their separation should be less than this threshold distance.
The blank region in the upper-right part of the figure corresponds to the conditions where Rydberg atoms undergo direct field ionization. The lower-left section represents a parameter range where our model is not applicable since the interatomic distance is too small for a classical treatment}
\end{figure}

To conclude, we report the first measurement of the Penning ionization rate of cold Rydberg atoms in the presence of a static electric field and find that it exceeds previous theoretical predictions by several orders of magnitude.
We demonstrate that this enhancement arises from two mechanisms induced by the static electric field together with the DDI between Rydberg atoms.

In studies related to quantum information or quantum simulation, relatively weak electric fields are often sufficient to induce a sensitive response in Rydberg atoms. 
However, a strong electric field is also useful in some cases, where the dipole-dipole interaction can be finely tuned to manipulate the Rydberg blockade effect~\cite{PhysRevLett.125.103602}. 
Our results indicate that under such strong electric field conditions, Penning ionization may lead to substantial Rydberg atom loss, posing a significant threat to system stability and coherence. 
Moreover, a strong electric field can also induce double ionization processes between Rydberg atoms. As illustrated in Fig.~\ref{fig:4}. Our model predicts that double ionization of n=100 Rydberg atom-pair occurs even if the electric field decrease to 2 V/cm.

{\it Acknowledgments} This work was supported by the National Key R$\&$D Program of China (Grant No. 2022YFA1602503) and the National Natural Science Foundation of China (Grant No. U2430208).

\nocite{*}

\bibliography{apssamp}

\end{document}